\documentclass[osajnl,twocolumn,showpacs,superscriptaddress,10pt]{revtex4-1}

\usepackage{amssymb}
\usepackage{amsmath}
\usepackage{graphicx}
\usepackage{dcolumn}
\usepackage{bm}
\usepackage{color}

\input{tcilatex}
\begin{document}

\title{Switching between positive and negative group delay of the optical pulse reflection
from layer structures with a Graphene sheet}
\author{Lin Wang}
\affiliation{Department of Physics, Zhejiang University, Hangzhou 310027, China}
\affiliation{Institute for Quantum Science and Engineering (IQSE) and Department of
Physics and Astronomy, Texas A$\&$M University, College Station, Texas
77843-4242, USA}
\author{Li-Gang Wang}
\email{sxwlg@yahoo.com}
\affiliation{Department of Physics, Zhejiang University, Hangzhou 310027, China}
\author{M. Suhail Zubairy}
\affiliation{Institute for Quantum Science and Engineering (IQSE) and Department of
Physics and Astronomy, Texas A$\&$M University, College Station, Texas
77843-4242, USA}

\begin{abstract}
In this paper, we investigate the propagation of the light pulse reflected
from the layer system with a graphene layer. We show a tunable
transition between positive and negative group delay of the optical pulse reflection in such a layered system
controlled by the properties of the graphene layer, and reveal two
mechanisms to control the propagation properties of the light reflected from
such systems. It is demonstrated that the reflected group delays are greatly
tunable from positive and negative values in both mechanisms of resonances and the
excitations of the surface plasmon resonances, which are
adjusted by tuning the Fermi energy and temperature of the graphene layer. Our results are helpful to control the pulse propagations and are useful for
design of graphene-based optical devices.
\end{abstract}

\date{\today}
\maketitle




\section{Introduction}

\textit{\ }Controlling the group delay (or group velocity) of a light pulse
has been extensively studied in both theories and experiments for many
years \cite{Milonni2005,Chiao1997,Brillouin1960,Khurgin2009}. When group velocity is much smaller than the speed of light $c$ in
vacuum, it often refers to slow light with a large value of group delay. The large positive group delay has potential applications in realizing optical delay lines \cite{Khurgin2009} and  a high-efficiency memory for optical pulses \cite{Novikova2012}%
. The approaches for realizing this delay have been widely investigated in various circumstances, such as ensembles of warm atoms (i.e. Rb and Cs) in vapor cells \cite{Novikova2012}, photonic crystal waveguides \cite%
{Kurt2013,Monat2010,Schulz2010}, and multiple quantum wells \cite{Yan2013}. Meanwhile, when group velocity is possible to be much larger than $c$ or
even becomes negative, it may be called as fast light with a very small or
negative group delay. It should be emphasized that this group delay do not violate causality or special relativity because
group velocity is not regarded as a signal velocity \cite{Chiao1997}. It has been observed in different systems including
left-handed media \cite{Woodley2004}, the samples of GaP:N \cite{ChuandWong1982}, quantum wells \cite{Vetter2001},\ double-Lorentzian fiber grating \cite{Longhi2002}, weakly absorbing slabs \cite{Wang2006},
atomic media \cite{Akulshin2010}, and ruby \cite{Gao2010}. Furthermore, in some system or media, such as fibers \cite{Arrieta-Yanez2010,Thevenaz2008}, semiconductor waveguides \cite{Mork2010,Mork2009}, nonlinear wave-mixing processes \cite{Bortolozzo2010}, solids at room temperature \cite{Zhang2008,Bigelow2004}, and gain slabs \cite{Wang2014}, the two types of group delay can been demonstrated simultaneously. Recently, with the emergence of new materials, controlling the group delay in other structures and media are
receiving more and more attention.

Graphene, a one-atom-thick allotrope of carbon, is focused
extensively in materials science and condensed-matter physics \cite%
{Bonaccorso2010, Geim2007}. It has linear dispersion relation of electronic
states characterized by conical and valence bands joined together at Fermi
level (the so called Dirac point), and the Fermi level can be controlled by
application of external or magnetic fields \cite%
{Castro2009,Christensen2012,Gusynin20061,Gusynin20062,Falkovsky2007,Vakil2011,Zi2013}%
. Due to the tunability, graphene shows many potential applications in
graphene-based nano-electronic and opto-electronic devices \cite%
{Novoselov2004,Castro2009}. Several research papers have been focused on the
manipulation of the light propagation in the system with graphene by
controlling the conductivity of graphene. In 2015, Hao et al. revealed that
graphene exhibits much stronger slow light capability than other materials
\cite{Hao2015}, and a large delay-bandwidth product has been obtained in
graphene-based waveguide. Meanwhile, Lu et al. designed a kind of plasmonic
structure consisting of a monolayer graphene to slow down and trap the light
in the mid-infrared region \cite{Lu2015}. Shi et al. found that the
plasmonic modes in the graphene nanostructure can be confined to a spacial
size that is hundreds of times smaller than their corresponding wavelength
in vacuum \cite{Shi2013}. Li et al. investigated the graphene ribbon
waveguide and achieved an outstanding plasmonically induced transparency
window with a group time up to 0.28ps \cite{Li2015}. On the other hand, in
2014 \cite{Jiang 2014}, Jiang et al. investigated the negative group delay
of the TE-polarized beam reflected from a Fabry-Perot cavity with the
insertion of the graphene. Next year, they found that a fast pulse
reflection can take place from the graphene covered lossless dielectric slab
\cite{Jiang 2015}. It is possible to realize the large positive and negative group delay in the same structure with the help of the graphene layer. In
addition, the Otto configuration combined with graphene have been
investigated, and the transverse magnetic surface plasmons
\cite{Mendieta 2014}, perfect terahertz absorption \cite{LYJiang 2016}, and
surface modes of transverse electric polarization \cite{Mason 2014,Mendieta
2015,Menabde 2016} have been proved and studied in succession in such a
configuration.

Motivated by these studies, we have theoretically considered the light
reflected from the layer configuration that incorporates graphene. Two
mechanisms that realize the group delay of the reflected pulse to be control
from positive to negative values, or vise verse, are demonstrated. The first is
related to resonances occurred when the incident angle is smaller than the
total internal reflection angle. The second one is related to the excitation
of surface plasmon happened when the incident angle is larger than the total
internal reflection angle. Both of cases lead to a distinct
variation of the group delays of the reflected light pulse, and it can be
controlled by adjusting the Fermi energy and the temperature of the graphene
sheet. Moreover, the structural parameters such as the position of the
graphene layer and angle of incidence can also provide an effective method
to control the reflected light pulse. Our results may have potential
application in graphene-based optical technologies and information
processing.

\begin{figure}[!tbp]
\includegraphics[width=8.2cm]{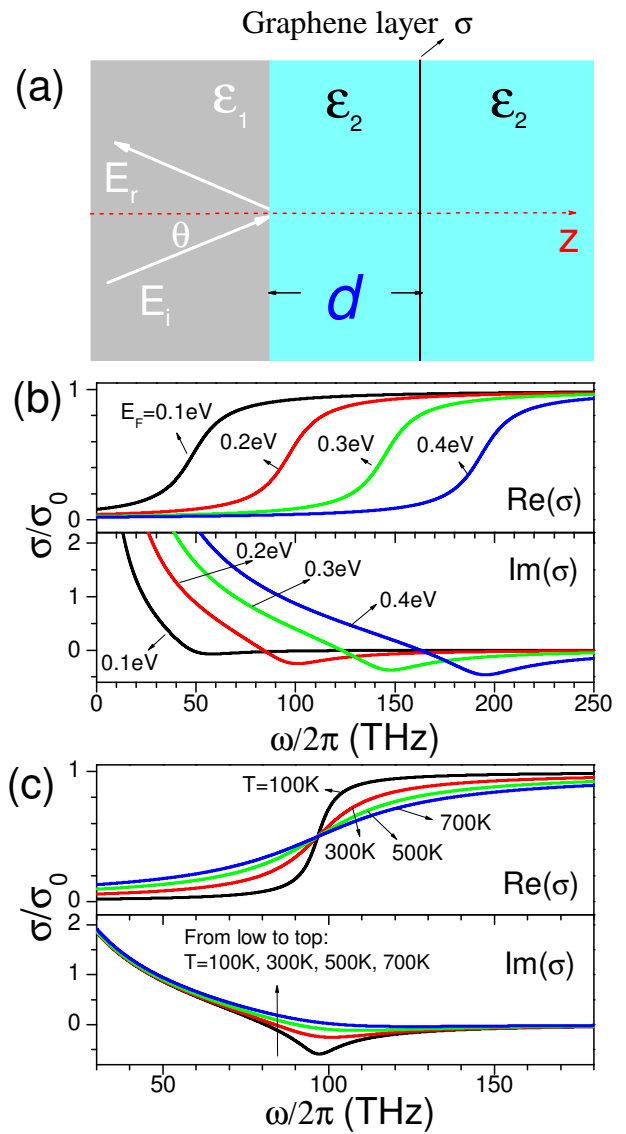}
\caption{(color online) (a) Schematic diagram of the layered-structure with
graphene. Here $\protect\varepsilon _{1}$ is the permittivity
of medium 1, and $\protect\varepsilon _{2}$ is the
permittivity of medium 2 (air or vacuum). In addition, it requires $\protect%
\varepsilon _{1}>\protect\varepsilon _{2}$, so the total internal reflection
can occur when the angle of incidence is larger than the critical angle. $%
\protect\theta $ and $d$ denote the incident angle and gap between the medium 1 and graphene, respectively, and $\protect\sigma $ represents the
conductivity of graphene. The effects of (b) the Fermi energy $E_{F}$ and
(c) temperature $T$ on the optical conductivities of the graphene sheet,
with $T=300$K and $E_{F}=0.2$eV, respectively. }
\label{Fig 1}
\end{figure}

\section{Model and Calculation}

Let a light pulse with a carrier frequency $\omega $ inside the
dielectric medium $\varepsilon _{1}$ be incident on the slab system
containing a graphene layer. As shown in Fig. 1(a), the graphene layer is
placed inside the dielectric medium $\varepsilon _{2}$ with a distance $d$
to the interface between dielectric media $\varepsilon _{1}$ and $%
\varepsilon _{2}$, and $\theta $ is the incident angle of light. Here it should be emphasized that since $\varepsilon_{1}>\varepsilon _{2}$, when the total internal reflection occurs, in this sense, the structure can be seen as Otto configuration similar to that in
Ref. \cite{Bludov2013}. The surface conductivity $\sigma $ of graphene is
usually given by Kubo formula. In the low temperature limit $k_{B}T\ll E_{F}$
and the zero collision rate ($\tau ^{-1}=0$), the surface conductivity can
be expressed as \cite{Falkovsky2007} $\sigma =\sigma _{\text{Intra}}+\sigma
_{\text{Inter}}$, where $\sigma _{\text{Intra}}=\frac{i8\sigma _{0}k_{B}T}{%
\pi \hbar \omega }\ln \left( 2\cosh \frac{E_{F}}{2k_{B}T}\right) $ is the
intraband electron transition contribution, and $\sigma _{\text{Inter}}=%
\frac{\sigma _{0}}{2}+\frac{\sigma _{0}}{\pi }\arctan (\frac{\hbar \omega
-2E_{F}}{2k_{B}T})-\frac{i\sigma _{0}}{2\pi }\ln [\frac{\left( \hbar \omega
+2E_{F}\right) ^{2}}{\left( \hbar \omega -2E_{F}\right) ^{2}+\left(
2k_{B}T\right) ^{2}}]$ is the interband electron transition contribution.
Here $\sigma _{0}=\frac{e^{2}}{4\hbar }$, $e$ is the charge of an electron, $%
\hbar $ is the reduced Plank's constant, $k_{B}$ is the Boltzmann constant, $%
T$ denotes the temperature, and $E_{F}$ is the Fermi energy which may be
electrically adjusted by the applied gate voltage. Clearly the real part of $%
\sigma $ is determined by $\sigma _{\text{Inter}}$ while its imaginary part
is determined by both $\sigma _{\text{Intra}}$ and $\sigma _{\text{Inter}}$.
From Figs. 1(b) and 1(c), it is seen that the optical conductivity of
graphene with frequency is shifted when the Fermi energy increases. In
addition, the optical conductivity of graphene is also affected by
temperature. Meanwhile, it is also noted that the real part of $\sigma $
(related to the absorption) changes dramatically near the energy of light
frequency close to $2E_{F}$. With these properties, it is expected that the
propagation properties of an optical structure containing graphene may be
controlled by both these external and structural parameters.

Generally speaking, there are two methods in treating graphene
as an element of a layered structure for optical applications. On one hand,
the graphene can be considered as a zero-thickness layer characterized by
the two dimensional conductivity \cite{Zi2013,Jiang 2014,Jiang 2015,Mendieta 2014,LYJiang 2016}. On the other hand, it can be seen as a
thin layer with effective dielectric constant \cite{Othman2013,EI-Naggar2015}. For our purposes, using one
of these methods is only for theoretical convenience and the same results
are expected by another method. Here we treat graphene
as a zero-thickness layer. According to the Maxwell's equation and the
boundary conditions, the reflection coefficient $r$ for TM-polarization is
given by%
\begin{equation}
r=\frac{r_{_{12}}+r_{_{\sigma }}e^{i2k_{2z}d}}{1+r_{_{12}}r_{_{\sigma
}}e^{i2k_{2z}d}},  \label{reflection}
\end{equation}%
where $r_{_{12}}=\frac{\varepsilon _{2}k_{1z}-\varepsilon _{1}k_{2z}}{%
\varepsilon _{2}k_{1z}+\varepsilon _{1}k_{2z}}$ is Fresnel reflection
coefficient between $\varepsilon _{1}$ and $\varepsilon _{2}$, $r_{_{\sigma
}}=\frac{\sigma k_{2z}}{2\varepsilon _{0}\varepsilon _{2}\omega +\sigma
k_{2z}}$ is the Fresnel reflection coefficient for the interface of a
graphene layer inside $\varepsilon _{2}$, $\varepsilon _{0}$ is the
permittivity of vacuum, and $k_{jz}=\left( k_{0}^{2}\varepsilon
_{j}-k_{y}^{2}\right) ^{1/2}$ $(j=1,2)$ is the $z$ component of the wave
vector inside the $j$th medium with $k_{y}=k_{0}\sqrt{\varepsilon _{1}}\sin
\theta $ and $k_{0}=\frac{\omega }{c}$. For the
narrow-spectrum incident pulse, i.e., $\Delta \omega \ll \omega $ ($\Delta
\omega $ is the spectrum width), the group delay of the reflected pulse can
be calculated by \cite{Sanchez2010}
\begin{equation}
\tau _{r}=\frac{d\phi _{r}\left( \omega \right) }{d\omega }\text{,}
\label{group delay}
\end{equation}%
where $\phi _{r}$ is the phase of the reflection coefficient. In our calculation, without loss of
generality, we assume $\varepsilon _{1}=2.25$ and $%
\varepsilon _{2}=1$ for vacuum (or air). This means that the Brewster angle $%
\theta _{B}$ of the system is $\theta _{B}=33.69%
{{}^\circ}%
$, and the critical angle $\theta _{c}$ of total reflection is $\theta
_{c}=41.8%
{{}^\circ}%
$.
\begin{figure}[tbp]
\includegraphics[width=8.2cm]{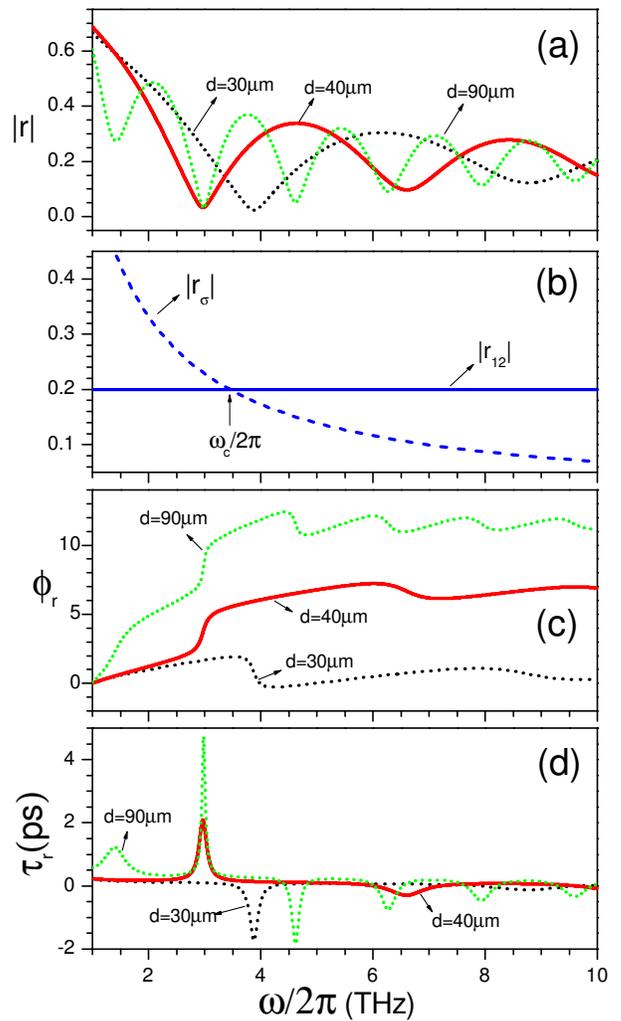}
\caption{(color online) Dependence of (a) $|r|$, (b) $|r_{12}|$ and $|r_{%
\protect\sigma }|$, (c) $\protect\phi _{r}$, and (d) $\protect\tau _{r}$ on
the frequency for normal incidence. Note that the dot, solid, and short
dashed lines, respectively, denote the cases of $d=30\protect\mu $m, $d=40%
\protect\mu $m and $d=90\protect\mu $m in Fig. 2(a), 2(c), and 2(d). The
other parameters are $E_{F}=0.2$eV and $T=300$K.}
\label{Fig 2}
\end{figure}

\section{Numerical Results and Discussions}

First we consider the case of normal incidence, and the results presented
here can be extended to the angles of incidence $\theta <\theta _{c}$. In
Fig. 2, we show the typical properties of light pulse reflected from the
layered system containing a graphene layer for different values of $d$ ($%
d=30\mu $m, 40$\mu $m and 90$\mu $m) in the case of normal incidence. It is
clear that the reflection dips in Fig. 2(a) are observed as a result of the
structural resonance. For convenience, we use $\omega _{R,j}$ to denote the
resonant frequencies with $j=1,2,3\cdots $. In Fig. 2(b), we compare the
magnitudes of Fresnel reflection coefficients $\left\vert r_{12}\right\vert $
and $\left\vert r_{\sigma }\right\vert $. In our calculation, $|r_{12}|=0.2$
for normal incidence, whereas $|r_{\sigma }|$ decreases with the increase of
frequency. Thus there is a critical frequency $\omega _{c}$ which makes $%
|r_{12}(\omega _{c})|=|r_{\sigma }(\omega _{c})|$, see Fig. 2(b). From Fig.
2, when $|r_{12}\left( \omega _{R,j}\right) |>|r_{\sigma }\left( \omega
_{R,j}\right) |$, the changes of phase with frequency for these resonances
are abnormally dispersive and the corresponding group delays are negative.
When $|r_{12}\left( \omega _{R,j}\right) |<|r_{\sigma }\left( \omega
_{R,j}\right) |$, the situation is totally reversed. For examples, in the
case of $d=30\mu $m, all resonances are in the region of $|r_{12}\left(
\omega _{R,j}\right) |>|r_{\sigma }\left( \omega _{R,j}\right) |$, therefore
the group delays of the reflected light pulse are negative near resonances.
In the case of $d=40\mu $m, one of resonances moves to the region of $%
|r_{12}|<|r_{\sigma }|$, thus the corresponding group delay becomes
positive. As $d$ increases, there are more numbers of resonances moving into
the region of $|r_{12}|<|r_{\sigma }|$; for instance, there are two
resonances satisfying this condition in the plots for $d=90\mu $m. For a
fixed graphene-based layered structure, $\varepsilon _{1}$, $\varepsilon
_{2} $ and $d$ are usually fixed and cannot be changed. However, it is
expected that we can change $r_{\sigma }$ by controlling the Fermi energy
and temperature, thus the propagation of light reflection is automatically
manipulated.

\begin{figure}[!tbp]
\includegraphics[width=8cm]{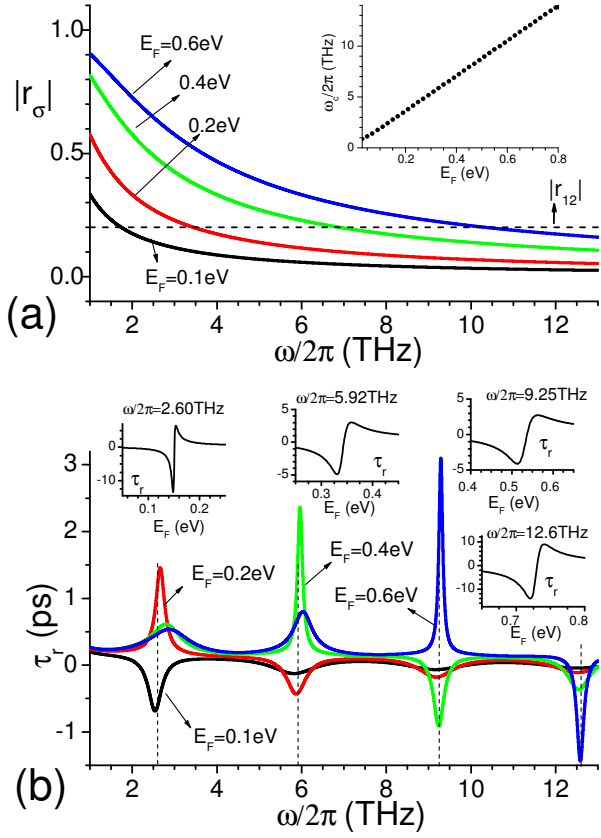}
\caption{(color online) Effects of the Fermi energy $E_{F}$ on (a) $%
\left\vert r_{\protect\sigma }\right\vert $ and (b) group delays $\protect%
\tau _{r}$. Inset in (a) shows the dependence of the critical frequency $%
\protect\omega _{c}$ on the value of $E_{F}$. Insets in (b) show the detail
dependence of $\protect\tau _{r}$ on $E_{F}$ at frequencies $\protect\omega %
/2\protect\pi =2.60$THz, $5.92$THz, $9.25$THz, and $12.60$THz near each
resonance, indicated by the vertical dash lines from left to right. The
other parameters are $\protect\theta =0^{\circ }$, $d=45\protect\mu $m, and $%
T=300$K.}
\label{Fig 3}
\end{figure}

From the above discussion on Fig. 2, it is observed that the critical
frequency $\omega _{c}$ at which the condition of $|r_{12}(\omega
_{c})|=|r_{\sigma }(\omega _{c})|$ holds is very important for implementing
the transition of phases and group delays near the corresponding resonances.
Since $r_{12}$ is a constant, one can change $r_{\sigma }$ by adjusting the
Fermi energy and temperature. In Fig. 3(a), we show that an increase in the
value of $E_{F}$ greatly shifts the curve for $\left\vert r_{\sigma
}\right\vert $ to the higher-frequency region. Thus the value of $\omega
_{c} $ in the system increases linearly with $E_{F}$, see the inset in Fig
3(a). Therefore, the reflected group delays near the frequencies of
resonances change their signs as $E_{F}$ increases in Fig. 3(b), and the
light pulses at different carrier frequencies can be manipulated from negative
to positive group delay reflection by simply adjusting the value $E_{F}$, see the inset in
Fig. 3(b).

\begin{figure}[tbp]
\includegraphics[width=8cm]{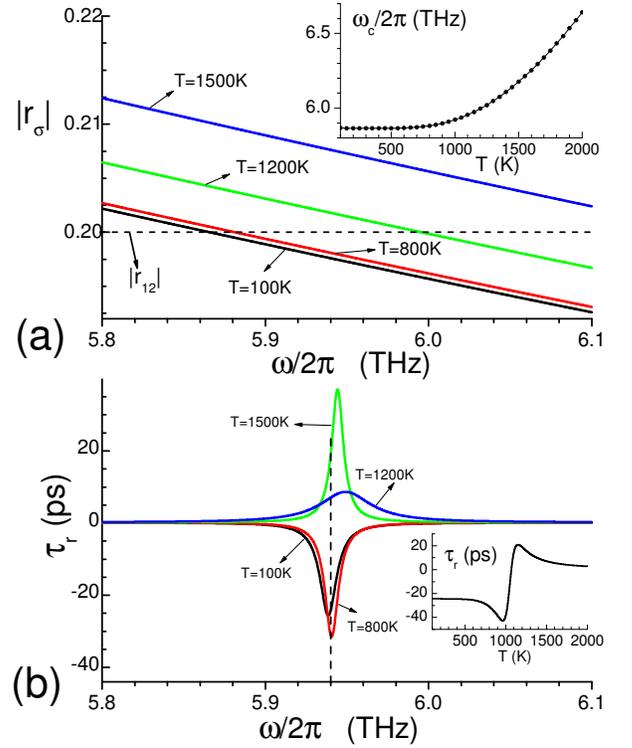}
\caption{(color online) Effects of temperature $T$ on the values of (a) $|r_{%
\protect\sigma }|$ and (b) $\protect\tau _{r}$. Here we take $T=100$K, $800$%
K, $1200$K, and $1500$K as examples. Inset in (a) shows the dependence of
the critical frequency $\protect\omega _{c}$ on temperatures $T$. Inset in
(b) displays the behavior of the group delay as a function of $T$ at
frequency $\protect\omega /2\protect\pi =5.94$THz, which is denoted by the
vertical dash line in (b). Other parameters are $\protect\theta =0^{\circ }$%
, $d=45\protect\mu $m, and $E_{F}=0.34$eV.}
\label{Fig 4}
\end{figure}

The effect of temperature on light reflection in such systems is very subtle
because temperature only controls the transition part of conductivity $%
\sigma $ near the $2E_{F}$. From Fig. 1(c), it is seen that changing
temperature has more distinct effect on the real part of $\sigma $, which
represents the absorption of the graphene layer. Of course, there are
significant changes in the imaginary part of $\sigma $ near the frequencies
close to $2E_{F}$. In Fig. 4, we show the effect of the temperature $T$ on
the light reflection in this system. Similar to the case of Fig. 3, the
curve of $r_{\sigma }$ is shifted by changing $T$, but the critical value of
$\omega _{c}$ is not linear with $T$. The dependence of the reflected group
delays on $T$ is very sensitive when the resonance of the system happens
around the frequency satisfying the equation $|r_{12}(\omega
_{c})|=|r_{\sigma }(\omega _{c})|$, see Fig. 4(b). Another interesting
feature is that, at the extremely low temperature, the group delay of the
reflected light pulse shows the sharp change near $2E_{F}$.

\begin{figure}[!tbp]
\includegraphics[width=8cm]{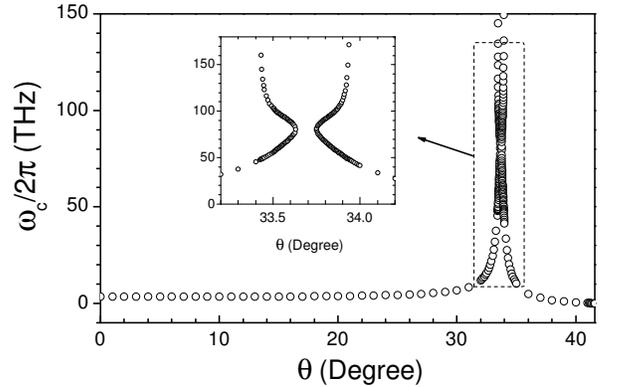}
\caption{The critical frequency $\protect\omega _{c}$ as a function of the
incident angle $\protect\theta $ for the case of $E_{F}=0.2$eV and $T=300$K.
Inset shows the detail near the Brewster angle $\protect\theta _{B}$.}
\label{Fig 5}
\end{figure}

Now we extend the above results into the cases of inclined incidence. When
the angle of incidence is $\theta <\theta _{c}$, the above results can be
readily obtained even for the cases of inclined incidence. Based on the
above condition $|r_{12}(\omega _{c})|=|r_{\sigma }(\omega _{c})|$, we
illustrate the dependence of the critical frequency $\omega _{c}$ on
incident angle in Fig. 5. From Fig. 5, we see that the critical frequency $%
\omega _{c}$ increases before the Brewster angle $\theta _{B}$. Near $\theta
_{B}$ there are two values of $\omega _{c}$ satisfying the above condition.
This is not surprising since $r_{\sigma }$ is not a monotonic function of
frequency. For certain angles, there are two regions satisfying $%
|r_{12}|<|r_{\sigma }|$. For example, as illustrated in Fig. 5, there may be
only one point of intersection between $|r_{12}|$ and $|r_{\sigma }|$ for
angles $\theta \in \lbrack 0,33.413%
{{}^\circ}%
)$ and $\theta \in (33.95%
{{}^\circ}%
,\theta _{c})$ or two points of such intersection for angles $\theta \in
(33.413%
{{}^\circ}%
,33.63%
{{}^\circ}%
)$ and $\theta \in (33.749%
{{}^\circ}%
,33.95%
{{}^\circ}%
)$, and even no point of such intersection for angles between $33.63%
{{}^\circ}%
$ and $33.749%
{{}^\circ}%
$. Therefore, by choosing the interesting frequency region and adjusting the
Fermi energy and temperature, we can naturally manipulate the group delays
in such systems.

Lastly, we discuss the case of total internal reflection (i.e., $\theta
>\theta _{c}$). In this situation, the layered structure may be seen as an
Otto structure. It is known that the effective dielectric constant of the
graphene layer can be given by $\varepsilon _{\sigma }=1+i\sigma /\omega
\varepsilon _{0}t_{g}$, where $t_{g}=0.5$nm is the thickness of graphene
sheet \cite{Koppens2011}. Thus Re$\left[ \varepsilon _{\sigma }\right] <0$
if Im$\left[ \sigma \right] >\omega \varepsilon _{0}t_{g}$. In this sense,
the graphene layer can be considered as the suitable alternative to a metal.
It is well known that the surface plasmon can be excited at a
metal-dielectric interface, and there is the counterintuitive dispersion
effect in the Otto configuration with metals \cite{Wang2016}. When the
graphene layer displays the properties of a metal, it is expected that the
light reflection in graphene-based Otto structures may also be manipulated
through the excitation of surface plasmon resonances.

\begin{figure}[!tbp]
\centering
\includegraphics[width=8cm]{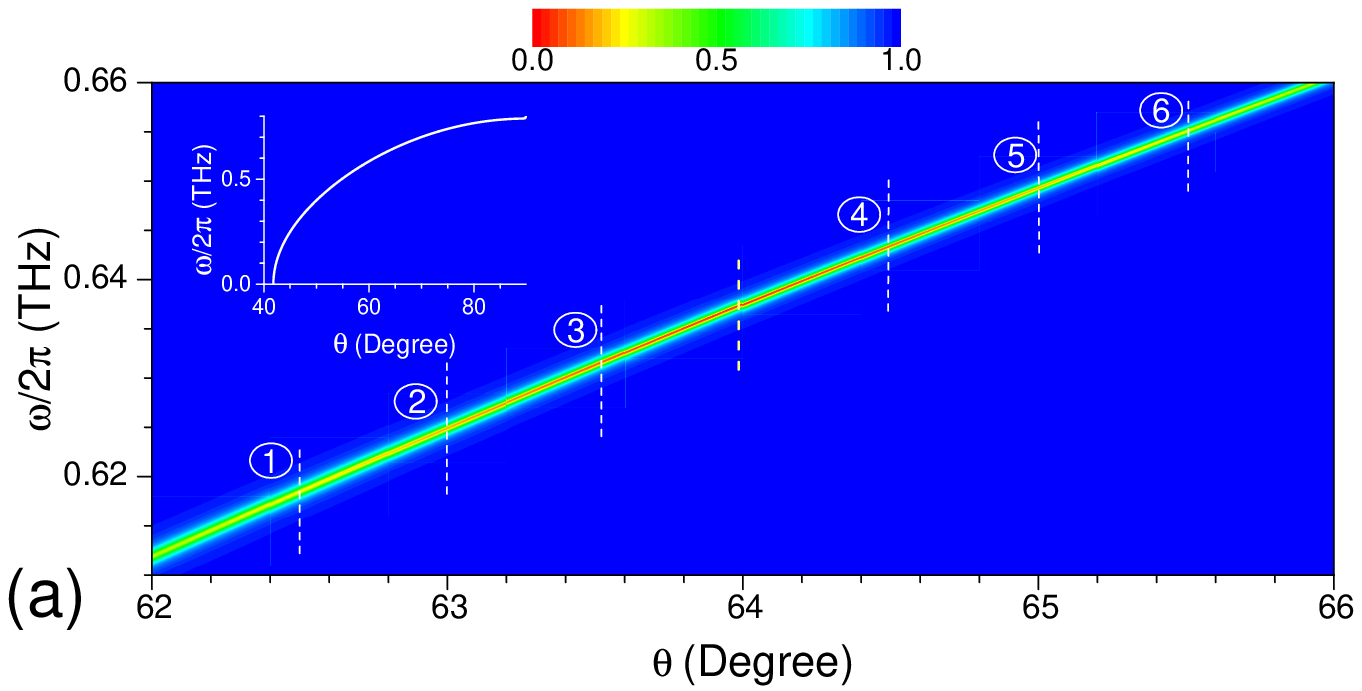} %
\includegraphics[width=8cm]{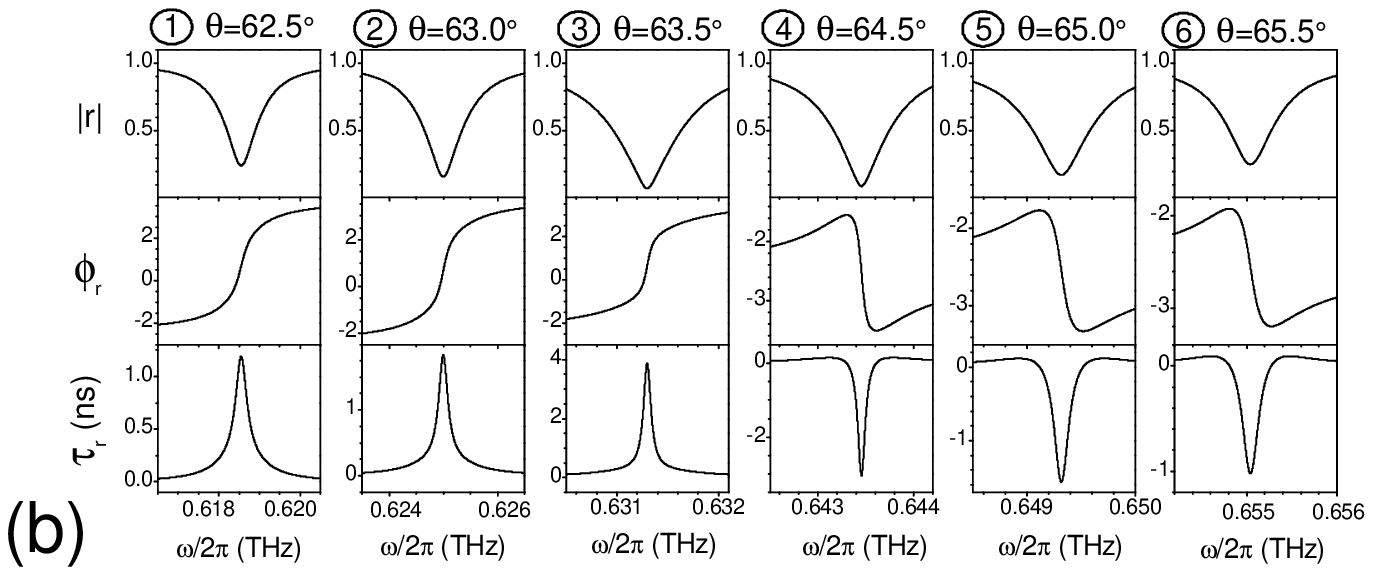}
\caption{(color online) (a) The existence of the optimal angle for the
excitation of the SPR. Inset shows the dispersion curve of the ideal surface
plasmon of a graphene layer in vacuum. Note that the yellow vertical dashed
line denotes the optimal angle $\protect\theta \approx 63.93^{\circ }$ which
is corresponding to the minimum of $|r|$. (b) The typical properties of the
reflection coefficient $|r|$, its phase shift $\protect\phi _{r}$, and the
corresponding group delay $\protect\tau _{r}$ at different incident angles $%
\protect\theta =62.5^{\circ }$, $63^{\circ }$, $63.5^{\circ }$, $64.5^{\circ
}$, $65^{\circ }$, and $65.5^{\circ }$. The other parameters are $E_{F}=0.2$%
eV, $T=300$K, and $d=300\protect\mu $m.}
\label{Fig 6}
\end{figure}

\begin{figure}[!tbp]
\includegraphics[width=8cm]{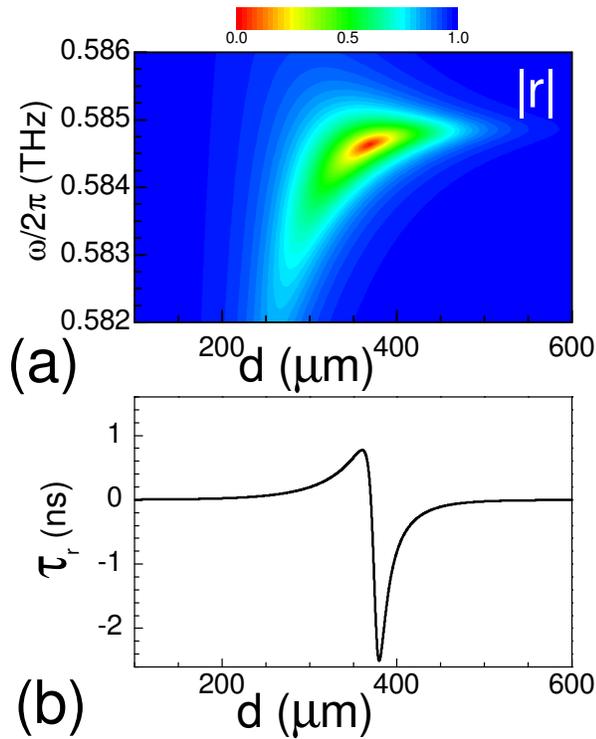}
\caption{(color online) (a) The existence of the optimal thickness for the
excitation of the SPR. (b) The behavior of the group delay $\protect\tau %
_{r} $ as a function of thickness $d$ at frequency $\protect\omega /2\protect%
\pi =0.5847$THz with $\protect\theta =60^{\circ }$. Other parameters are the
same in Fig. 6.}
\label{Fig 7}
\end{figure}

In Fig. 6(a), we show the typical excitation of surface plasmon resonance
(SPR) in such graphene-based Otto systems. There exists an optimal angle to
excite the SPR for a fixed value of thickness $d$. For example, in the case
of $d=300\mu $m, the optimal excitation of SPR is located at the angle
around $63.93%
{{}^\circ}%
$ denoted by the vertical dashed line in Fig. 6(a). This optimal angle
decreases along the dispersive curve of the ideal surface plasmon, see the
inset of Fig. 6(a), as the value of $d$ increases. In Fig. 6(b), we show the
physical properties of light reflection, its phase shift and group delay
before and after the optimal angle of the SPR. It is clear that the pulse
reflection suffers the transition from positive
to negative group delay. Similarly, there
exists an optimal thickness $d$ for a certain angle of incidence. As shown
in Fig. 7(a), in the case of $\theta =60%
{{}^\circ}%
$, the optimal thickness $d$ is around 368$\mu $m. Near the excitation
frequency of the SPR, it is clearly seen that the reflected group delay is
tunable by adjusting the thickness $d$, see Fig. 7(b).

\begin{figure}[!tbp]
\includegraphics[width=8cm]{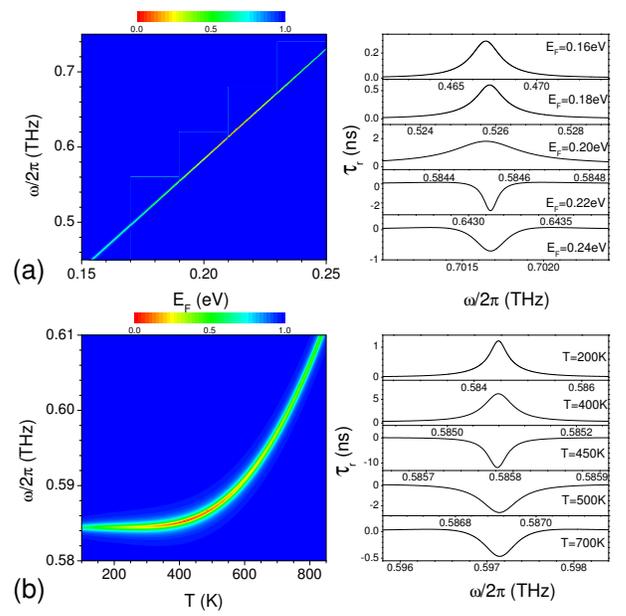}
\caption{(color online) (Left) Dependence of the function $|r|$ on (a) the
Fermi energy $E_{F}$ and (b) temperature $T$. (Right) The corresponding
reflected group delays as functions of frequencies under different values of
$E_{F}$ and $T$. Here $T=300$K in (a) and $E_{F}=0.2$eV in (b), and other
parameters are $\protect\theta =60^{\circ }$ and $d=350\protect\mu $m.}
\label{Fig 8}
\end{figure}

Furthermore, as discussed in the cases of resonance, the Fermi energy and
the temperature may also significantly affect on the properties of light
reflection in the cases of SPR excitations. Thus the reflected group delays
are tunable by controlling the values $E_{F}$ and $T$. In Fig. 8(a), we show
the linear relation of the SPR frequency vs the Fermi energy, and the group
delays show different behaviors as $E_{F}$ changes. There also exists a
transition value of $E_{F}$, which may lead to the changes of group delay
from positive to negative values near the SPR frequency, see the right part
of Fig. 8(a). It can be seen that the effect of $E_{F}$ on the SPR frequency
is similar to the previously discussed resonant situations. However only one
optimal $E_{F}$ exists when other parameters (like temperature, thickness $d$
and angle) are unchanged. There are similar effects of temperature $T$ on
the SPR frequency and group delays. Of course, the effect of temperature on
the SPR frequency is not linear, see Fig. 8(b).

\section{Summary}

We have presented the tunable transition effect of light pulse reflection in a layer system which contains a graphene
layer. It is shown that there are two mechanisms to control the properties
of light pulse reflection. When the angle of incidence is less than the
critical angle of total reflection, the reflected group delay can be greatly
tuned near the frequencies of resonances where the magnitudes of the Fresnel
reflection coefficients satisfy certain conditions. In the case of total
reflection, the optimal excitation of the SPR is the critical value to
adjust the properties of pulse reflection in such Otto systems. It is demonstrated that the reflected group delays are controllable and
tunable by adjusting the Fermi energy and temperature of the graphene sheet.
The reflected group delays can also manipulate via changing the structural
parameters including the position of the graphene layer and the angle of
incidence. These results are useful to understand the fundamental physics
underlying graphene-based light-matter interactions, and they may have
potential applications in graphene-based optical signal processing and
optical sensing.

\begin{acknowledgments}
This work is supported by the National Natural Science Foundation of China
(NSFC) (grants No. 11674284 and U1330203). It is also supported by the
Fundamental Research Funds for the Center Universities (No. 2017FZA3005).
This research is also supported by NPRP Grant No. 8-751-1-157 from the Qatar
National Research Fund. L. Wang was supported by the China Scholarship
Council (Grant No. 201606320146).
\end{acknowledgments}

\end{document}